\PassOptionsToPackage{x11names}{xcolor}
\documentclass[conference]{IEEEtran}
\IEEEoverridecommandlockouts
\usepackage{cite}
\usepackage{amsmath,amssymb,amsfonts}
\usepackage{algorithmic}
\usepackage{graphicx}
\usepackage{textcomp}
\usepackage{xcolor}
\usepackage{pifont}     %

\def\BibTeX{{\rm B\kern-.05em{\sc i\kern-.025em b}\kern-.08em
    T\kern-.1667em\lower.7ex\hbox{E}\kern-.125emX}}

\newcommand{\mb}[1]{\mathbf{#1}}

\newcommand{\bs}[1]{\boldsymbol{#1}}

\newcommand{\argmin}{\mathop{\rm arg~min}\limits}
\newcommand{\cmark}{\ding{51}}  %

\usepackage{comment}
\usepackage{amsmath,graphicx,algorithm,algorithmic,adjustbox,setspace}
\usepackage{svg}
\svgsetup{inkscapeexe=inkscape} 
\usepackage{subcaption,afterpage}
\usepackage{bm}
\usepackage{float}
\usepackage{placeins} %
\usepackage{siunitx}
\usepackage{booktabs} 

\let\OLDthebibliography\thebibliography
\renewcommand\thebibliography[1]{
  \OLDthebibliography{#1}
  \setlength{\parskip}{2pt}
  \setlength{\itemsep}{2pt plus 1.5ex}
}

\usepackage{ifthen}
\newboolean{showchanges}
\setboolean{showchanges}{false} %

\usepackage[x11names]{xcolor}
\newcommand{\revise}[1]{%
  \ifthenelse{\boolean{showchanges}}%
    {\textcolor{Red3}{#1}}%
    {#1}%
}

\begin{document}

\title{Exploring Disentangled Neural Speech Codecs\\from Self-Supervised Representations}

\author{
    Ryo Aihara\IEEEauthorrefmark{1}\IEEEauthorrefmark{2},
    Yoshiki Masuyama\IEEEauthorrefmark{2},
    Gordon Wichern\IEEEauthorrefmark{2},
    Fran\c{c}ois G.\ Germain\IEEEauthorrefmark{2},
    Jonathan Le Roux\IEEEauthorrefmark{2}
    \vspace{.7\baselineskip}
    \IEEEauthorblockA{\\
        \IEEEauthorrefmark{1}\textit{Information Technology R\&D Center},
        \textit{Mitsubishi Electric Corporation},
        Kanagawa, Japan, \\
        Aihara.Ryo@dx.MitsubishiElectric.co.jp
    }
    \IEEEauthorblockA{
        \IEEEauthorrefmark{2}\textit{Mitsubishi Electric Research Laboratories (MERL)},
        MA, USA, \\
        \{masuyama, germain, wichern, leroux\}@merl.com
    }
}

\maketitle

\begin{abstract}
    Neural audio codecs (NACs), which use neural networks to generate compact audio representations, have garnered interest for their applicability to many downstream tasks---especially quantized codecs due to their compatibility with large language models. However, unlike text, speech conveys not only linguistic content but also rich paralinguistic features. Encoding these elements in an entangled fashion may be suboptimal, as it limits flexibility. For instance, voice conversion (VC) aims to convert speaker characteristics while preserving the original linguistic content, which requires a disentangled representation. Inspired by VC methods utilizing $k$-means quantization with self-supervised features to disentangle phonetic information, we develop a discrete NAC capable of structured disentanglement. Experimental evaluations show that our approach achieves reconstruction performance on par with conventional NACs that do not explicitly perform disentanglement, while also matching the effectiveness of conventional VC techniques.
\end{abstract}


\section{Introduction}
\label{sec:intro}
The recent success of large language models (LLMs) has \revise{renewed} interest in neural audio codecs (NACs) 
within the speech and audio research community. 
NACs \revise{have found use} not only as tokenizers for audio and speech language models such as AudioLM~\cite{borsos2023audiolm},
but also to produce tokens as outputs in audio-based dialogue systems~\cite{kyutai2024moshi}.
NACs exist in both continuous and discrete flavors, and which one is better is an open research question.
While continuous representations offer strong reconstruction capabilities~\cite{mousavi2024dasb}, 
discrete audio tokens are often preferred due to their compatibility with LLMs, 
which operate on discrete text tokens~\cite{kyutai2024moshi}.

End-to-end discrete NACs have been extensively studied. They typically leverage a generative adversarial network (GAN) framework 
to train a convolutional encoder for processing audio waveforms, a quantizer employing residual vector quantization (RVQ), 
and a deconvolutional decoder for waveform reconstruction~\cite{zeghidour2021soundstream,defossez2022highfi, kumar2023highfidelity}.

Another key direction in the development of NACs is disentanglement.
Several studies have leveraged distillation from self-supervised learning (SSL) models or 
integrated SSL to capture ``semantic'' information, 
leading to improved ASR performance with these tokens~\cite{zhang2024speechtokenizer, huang2024repcodec, ye2024codec}.
Considering ASR and other potential downstream tasks for speech codecs, it is desirable to develop disentangled codecs.

Speech information can be categorized into three types: linguistic (lexical, semantic), paralinguistic (intentional, stylistic), and non-linguistic (physical, emotional)~\cite{fujisaki2003prosody}.
This suggests that a speech codec be disentangled into these three categories to more effectively capture the distinct aspects of speech.
Voice conversion (VC), which aims to modify speaker characteristics while preserving the original linguistic content, also necessitates a disentangled representation. 
In recent years, SSL has been increasingly employed to enhance this process.
In~\cite{baas2023voice}, it was shown that the middle layer of WavLM~\cite{chen2022wavlm} is strongly correlated with speaker identity 
while retaining prosody and phonetic information,
enabling speaker VC by swapping the corresponding WavLM representations between two speakers.
This approach has also been applied to text-to-speech (TTS) systems~\cite{hajal2025knn}.
Similarly, Sim et al.~\cite{sim2024skqvc} \revise{proposed} a one-shot VC method that also leverages the middle layer of WavLM, 
employing $k$-means quantization.

Inspired by these methods, this study explores a speech codec that disentangles phonetic, prosodic, and speaker characteristics, which correspond to linguistic, paralinguistic, and non-linguistic information. 
To evaluate the proposed approach, we conduct both reconstruction and one-shot VC experiments.
The results are compared with a non-disentangled codec and conventional VC methods.

\revise{The contributions of this paper can be summarized as follows. 
First, we explore a fully discrete speech codec that disentangles phonetic, prosodic, and speaker information, without any supervision from phonetic labels or fundamental frequency ($F_0$). 
We also demonstrate that our disentangled codec achieves one-shot voice conversion performance comparable to conventional methods, using only a single utterance from the target speaker. Experimental results indicate that discretizing speaker vectors tends to diminish speaker-specific information, requiring a sufficiently rich discrete representation to preserve speaker identity.}

\revise{
This approach is valuable on two levels: it enhances interpretability while also enabling disentangled control of a representation compatible with downstream applications like audio and speech language models (LMs).
}

\begin{table*}[t]
  \centering
  \caption{Comparison of related works on disentangled speech representations}
  \label{tab:paper_comparison}
  \begin{tabular}{
    l   %
    c   %
    c   %
    c   %
    c   %
  }
    \toprule
    & \textbf{Quantize} 
      & \textbf{Quantize} 
      & \textbf{Quantize} 
      & \textbf{No external} \\
    \textbf{Methods} 
      & \textbf{phoneme} 
      & \textbf{prosody} 
      & \textbf{speaker} 
      & \textbf{label} \\
    \midrule
    Phonetic Codecs \cite{zhang2024speechtokenizer, kyutai2024moshi, huang2024repcodec, ye2024codec, guo2024lscodec, qian2022contentvec, ren2024fewertoken, dellalibera2025focalcodecl, wang2021vqmivc, kanagawa2024knowledgs, sim2024skqvc, baas2023voice} & \cmark &       &       & \cmark \\
    Speech Resynthesis~\cite{polyak2021speechresynthesis} & \cmark & \cmark &  &       \\
    Phoneme-level Speech Codec~\cite{karapiperis2024investigating} & \cmark & \cmark &  &  \\
    FACodec~\cite{ju2024naturalspeech} & \cmark & \cmark & \cmark &  \\
    Ours & \cmark & \cmark & \cmark & \cmark \\
    \bottomrule
  \end{tabular}
\end{table*}
\section{Related works}
\label{sec:related}
\revise{
    \textbf{Neural audio codecs ---}  
        SoundStream~\cite{zeghidour2021soundstream} leveraged a GAN framework 
        to train a convolutional encoder for processing audio waveforms, RVQ, and a deconvolutional decoder
        for waveform reconstruction.
        EnCodec~\cite{defossez2022highfi} extended SoundStream by incorporating 
        a multi-scale discriminator and a loss-balancing mechanism to improve training stability.
        DAC~\cite{kumar2023highfidelity} refined RVQ through $L^2$-normalization and dimensionality reduction, serving as a strong baseline for subsequent advancements in neural audio coding.
        SNAC~\cite{siuzdak2024snac} extended DAC with muti-scale RVQ, leading to even more efficient compression.
        We employ DAC and SNAC as our baseline non-disentangled NACs.
}

\textbf{Disentangled speech representations ---}
    \revise{
        We summarized the comparison of some related works in Table~\ref{tab:paper_comparison}.
        Mimi~\cite{kyutai2024moshi}  and SpeechTokenizer~\cite{zhang2024speechtokenizer}
        introduced knowledge distillation from WavLM to capture phonetic information in codec feature space.
        Repcodec~\cite{huang2024repcodec} encodes and decodes directly from the SSL model, and
        X-Codec~\cite{ye2024codec} concatenates a phonetic codec with an acoustic codec, 
        resulting in enhanced ASR performance.
        However, the disentanglement of prosodic and speaker information is not investigated
        in these papers.
    }
    Polyak et al.~\cite{polyak2021speechresynthesis} proposed a speech resynthesis method 
    where phonetic, prosodic, and speaker information are estimated in parallel.
    However, their approach utilizes the estimated $F_0$
    from a vocoder as the target for prosodic quantization,
    which differs from our method. It also does not quantize the speaker vector.
    \revise{We assume that quantized speaker information is likely to be more compatible with downstream applications like speech LM.} 
    Ju et al.~\cite{ju2024naturalspeech} \revise{ proposed FACodec, which estimates three streams in parallel 
    for TTS, but their method requires phoneme labels for training, whereas our approach does not impose this 
    constraint. Additionally, they employ a gradient reversal layer (GRL) for disentanglement,
    which is known to be challenging to train effectively~\cite{series2020adversarial}.}
    Guo et al.~\cite{guo2024lscodec} applied speaker disentanglement in the codec feature space
    but \revise{their method} relies on the the waveform similarity overlap-add (WSOLA) technique, 
    which imposes significant computational costs.
    Similarly, Qian et al.~\cite{qian2022contentvec} introduced a speaker disentanglement method
    using contrastive learning between speakers,
    but their architecture does not incorporate quantization of prosodic information.
    Ren et al.~\cite{ren2024fewertoken} proposed a model that estimates a time-invariant code,
    and Della Libera et al.~\cite{dellalibera2025focalcodecl} introduced a \revise{NAC with a single codebook}, 
    which also disentangles speaker information.
    However, both approaches do not explicitly disentangle prosodic features from phonetic representations.
    Karapiperis et al.~\cite{karapiperis2024investigating} explored prosodic quantization,
    but their approach relies on externally provided phoneme sequences and speaker vectors.

\revise{
    \textbf{Voice conversion ---}
        VQ is utilized for VC to disentangle speaker information from other information~\cite{wang2021vqmivc}.
        SSL is also applied to VC to disentangle speaker information from phonetic information.
        Baas et al.~\cite{baas2023voice} proposed a simple but effective VC method using WavLM, 
        however, the conversion quality degrades when the number of target speaker utterances 
        is small during inference.
        Kanagawa et al.~\cite{kanagawa2024knowledgs} proposed to perform VC using discrete units from SSL, 
        however prosodic and speaker information is not discretized.
        SKQVC~\cite{sim2024skqvc} also uses discrete units from WavLM and we 
        employ this method as our baseline, where only phonetic information is discretized.
}

\begin{figure*}[t]
  \centering
  \begin{subfigure}{0.53\textwidth}
    \centering
    \includegraphics[width=.99\linewidth]{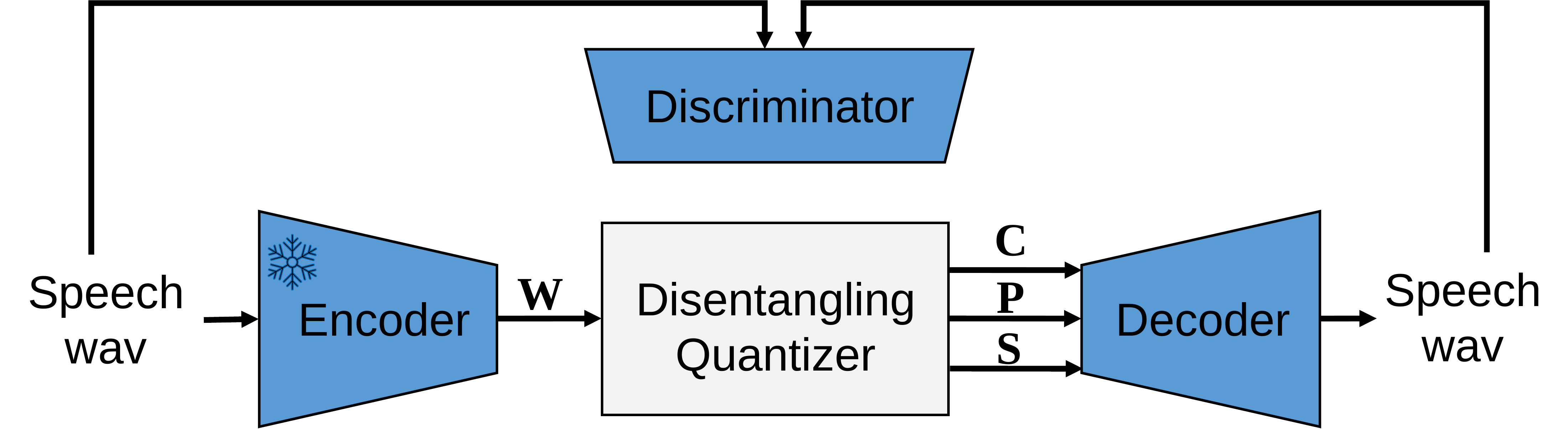}
    \vspace{-5mm}
    \subcaption{Overview of disentangled speech codec.}
    \vspace{+5pt}
    \label{fig:overview}
  \end{subfigure}
  \begin{subfigure}{0.53\textwidth}
    \centering
    \includegraphics[width=.99\linewidth]{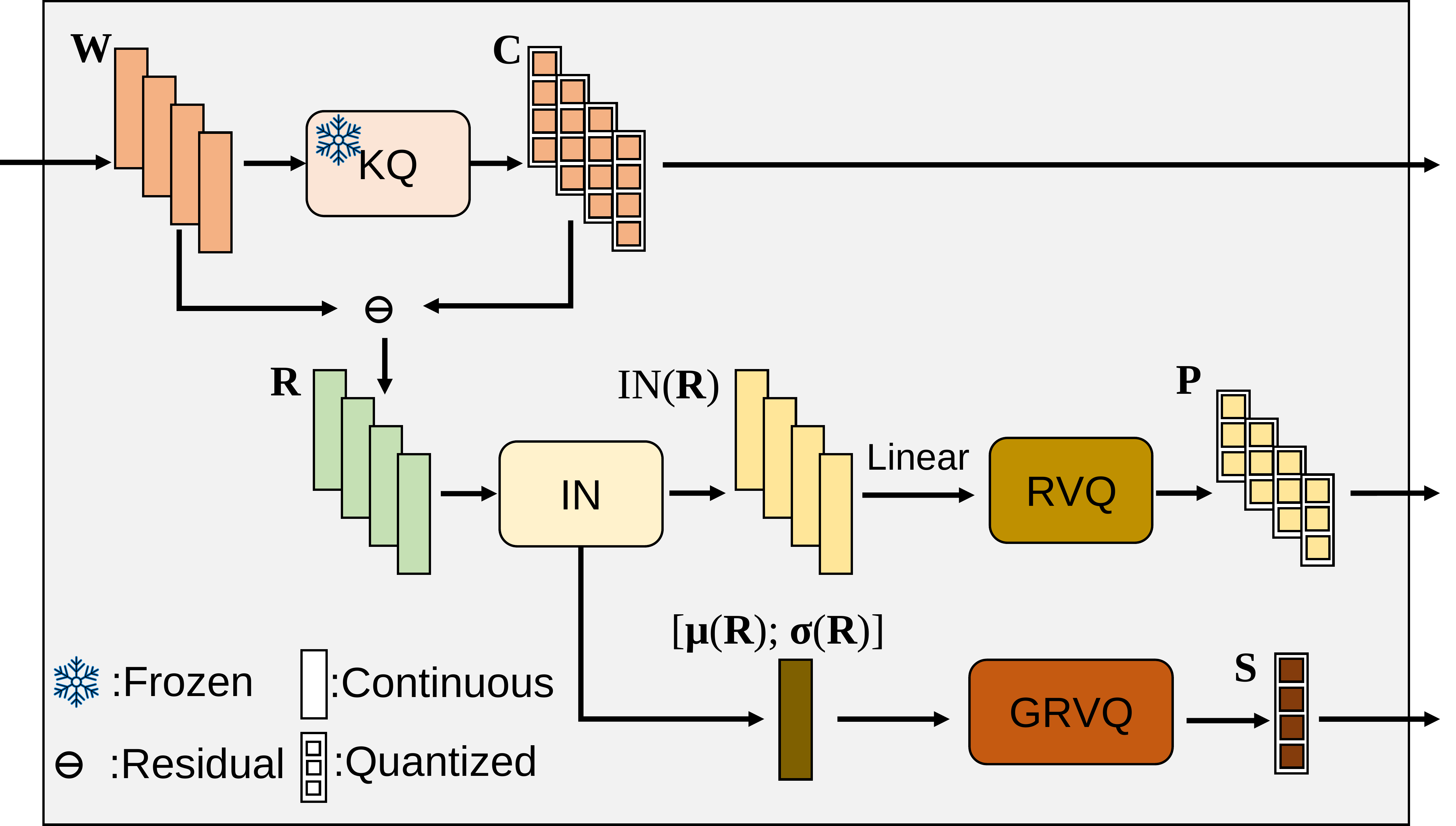}
    \subcaption{Disentangling quantizer.}
    \label{fig:Disentangling}
  \end{subfigure}
  \caption{Model architecture of disentangled speech codec.}
  \label{fig:main}
\end{figure*}

\section{Disentangled neural speech codec}
\label{sec:prop}
\subsection{Overview}
\label{subsec:overview}
Figure~\ref{fig:main} depicts the model architecture for our proposed disentangled speech codec, 
which is comprised of an encoder, disentangling quantizer, decoder, and discriminator. \revise{Here, the key distinction from conventional SKQVC lies in the use of a disentangling quantizer.}
We use WavLM-Large as the encoder, outputting its 6th-layer hidden vectors as representation. We denote them as $\mb{W} = \{\mb{w}_1, \mb{w}_2, \dots, \mb{w}_T\} \in \mathbb{R}^{D \times T}$,
where $D$ denotes the number of dimensions and $T$ denotes the number of frames.
We use a pre-trained WavLM~\cite{chen2022wavlm} and its parameters are frozen during the training of the codec.

The disentangling quantizer converts $\mb{W}$ into three components.
As shown in Fig.~\ref{fig:main},
$\mb{C} = \{\mb{c}_1, \mb{c}_2, \dots, \mb{c}_T\} \in \mathbb{R}^{D \times T}$,
$\mb{P} = \{\mb{p}_1, \mb{p}_2, \dots, \mb{p}_T\} \in \mathbb{R}^{F \times T}$, and
$\mb{S} \in \mathbb{R}^{2D}$
represent the quantized content vectors, quantized prosody vectors, and quantized speaker vector, respectively. $F$ denotes the dimension of the quantized prosody vectors. 
The decoder combines these components to reconstruct the speech waveform.
\revise{Note that the decoder of SKQVC accepts these three components but only the phonetic components are quantized while the other vectors remain continuous.}

The discriminator consists of a multi-period discriminator~\cite{kong2020hifigan}
and a complex multi-scale short-time Fourier transform (STFT) discriminator~\cite{jang2021univnet}.
The entire model is trained in a generative adversarial manner, as outlined in~\cite{kumar2023highfidelity}.

\subsection{Disentangling Quantizer}
\label{subsec:disentangling}
The quantized content vectors $\mb{C}$ are estimated from the representation $\mb{W}$ using VQ
as follows:
\begin{align}
    \mb{c}_t = \mb{q}_{\mkern4mu\hat{\mkern-4mu j}_t}, \quad \mkern5mu\hat{\mkern-5mu j}_t = \argmin_{j} \lVert \mb{w}_t - \mb{q}_j \lVert, \forall t,
\end{align}
where $\mb{Q} = \{\mb{q}_1, \mb{q}_2, \dots, \mb{q}_J\} \in \mathbb{R}^{D \times J}$ 
represents the codebook vectors and $J$ denotes the number of codes. 
\revise{We denote this operation as KQ in Fig.~\ref{fig:main}.}
We compute $\mb{Q}$ prior to training the quantizer, as the result of applying $k$-means clustering to the representations $\mb{W}$ obtained from a separate large dataset~\cite{mousavi2024how}. 
We then freeze $\mb{Q}$ for the rest of the training. 

Following the content quantization module KQ, we compute the residual $\mb{R}=\mb{W}-\mb{C} \in \mathbb{R}^{D \times T}$. %
Since the 6th layer of WavLM-Large has been shown to exhibit a high correlation with speaker identity and prosody compared to other layers~\cite{baas2023voice},
and the centroids of $k$-means of SSL hidden features are commonly assumed to capture speaker-independent phonetic information~\cite{hsu2021hubert},
the residual is likely to preserve substantial speaker-related and prosodic information.
Next, instance normalization (IN) is applied to $\mb{R}$ as follows:
\begin{align}
    \mathrm{IN(\mb{R})} = \frac{\mb{R} - \bs{\mu(R)}}{\bs{\sigma(R)}},\label{eq:IN}
\end{align}
where $\bs{\mu(R)}\in \mathbb{R}^{D}$  and $\bs{\sigma(R)} \in \mathbb{R}^{D}$ 
denote the mean and standard deviation of $\mb{R}$ along the time axis.
As a result, $\bs{\mu(R)}$ and $\bs{\sigma(R)}$ are time-invariant,
and their concatenated vector is treated as a pre-quantization speaker vector~\cite{chen2020againvc}. They are broadcasted appropriately to match the dimension of $\mb{R}$ in \eqref{eq:IN}.

Time-variant vectors $\mathrm{IN(\mb{R})} \in \mathbb{R}^{D \times T}$ are then assumed to contain prosodic information.
\revise{Because an information bottleneck is required} to minimize potential interference with phonetic content, 
these vectors are projected to an $F$-dimensional space by a linear transformation,
and quantized prosody vectors $\mb{P}$ are estimated by RVQ.
The prosodic quantization codebooks are estimated during training in the $L^2$-normalized feature space.

Finally, we use group-residual vector quantization (GRVQ)~\cite{yang2023hifi} to quantize the pre-quantization speaker vector,
and obtain the quantized speaker vector $\mb{S}$. 
In GRVQ, the input feature space is partitioned into several groups along the feature dimension axis, with RVQ performed on each group.
During the RVQ process for each group within GRVQ, %
codes are estimated in the $L^2$-normalized feature space
with the low-dimensional code lookup process as described in~\cite{kumar2023highfidelity}.

\subsection{Decoder}
\label{subsec:Decoder}
The quantized content vectors $\mb{C}$,
the quantized prosody vectors $\mb{P}$,
and the quantized speaker vector $\mb{S}$ 
are passed through the decoder.
The quantized content vectors and the quantized prosody vectors 
are concatenated along the dimension axis
and modulated by a feature-wise linear modulation (FiLM)~\cite{perez2018film} conditioned on the quantized speaker vector as follows:
\begin{align}
    \mathrm{FiLM}([\mb{C}; \mb{P}]|\mb{S}) &= f(\mb{S}) \odot [\mb{C}; \mb{P}] + h(\mb{S}),\label{eq:film}
\end{align}
where $f$ and $h$ can be \revise{trainable} arbitrary functions and %
$[\cdot; \cdot]$ and $\odot$
denote the concatenation operation along the dimension axis and the element-wise product, respectively.
In this paper, we employ a simple linear transformation for both functions.

The remainder of the decoder consists of up-sampling deconvolutional blocks 
that up-sample the modulated features to reconstruct the waveform. 
Following the approach presented in~\cite{siuzdak2024snac}, depth-wise deconvolution is applied to all deconvolutional layers, 
and a noise block is inserted after each up-sampling layer.

\section{Experiments}
\label{sec:experiments}
\begin{table*}[t]
    \caption{Reconstruction results for each method. Phonetic/prosodic/speaker indicate how each information is represented.}
    \label{table:reconstruction}
    \centering
     \sisetup{
    detect-weight, %
    mode=text, %
    tight-spacing=true,
    round-mode=places,
    round-precision=2,
    table-format=1.2,
    table-number-alignment=center
    }
    \begin{tabular}{lccc*{4}{S}}
        \toprule
        Model            & phonetic      & prosodic       & speaker    & {Mel-d $\downarrow$} & {STFT-d $\downarrow$} & {ViSQOL $\uparrow$} & {SV [\%] $\downarrow$} \\ 
        \cmidrule(lr){1-1}\cmidrule(lr){2-4}\cmidrule(lr){5-8}
        \revise{Vanilla} RVQ & \multicolumn{2}{c}{1.20 kb/s}  & -          & 0.82                 & 1.61                  & 4.01                & 3.01                   \\
        SNAC             & \multicolumn{2}{c}{1.05 kb/s}  & -          & 0.81                 & 1.60                  & 4.01                & 2.93                   \\
        SKQ              & 0.50 kb/s     & 1024 dim, 50Hz & 1024 dim/u & 1.01                 & 1.84                  & 3.97                & 2.36                   \\
        SKQ+$\sigma$     & 0.50 kb/s     & 1024 dim, 50Hz & 2048 dim/u & 1.04                 & 1.88                  & 3.93                & 2.48                   \\
        SKQ2+$\sigma$    & 0.50 kb/s     & 1.00 kb/s      & 2048 dim/u & 1.06                 & 1.89                  & 3.87                & 2.55                   \\
        SKQ3+$\sigma$    & 0.50 kb/s     & 1.00 kb/s      &  1.28 kb/u & 1.05                 & 1.88                  & 3.88                & 2.65                   \\ 
        \bottomrule
    \end{tabular}
\end{table*}
\begin{table*}[t]
    \caption{One-shot VC results for each method. ``c'' and ``q'' indicate continuous and quantized vectors, respectively.}
    \label{table:conversion}
    \centering
     \sisetup{
    detect-weight, %
    mode=text, %
    tight-spacing=true,
    round-mode=places,
    round-precision=2,
    table-format=1.2,
    table-number-alignment=center
    }
    \begin{tabular}{lccS[table-format=2.2]SSS}
        \toprule
        Model          & prosodic & speaker & {WER [\%] $\downarrow$} & {SV [\%]  $\downarrow$} & {UTMOSv2 $\uparrow$} & {$F_0$-PCC $\uparrow$} \\ 
        \cmidrule(lr){1-1}\cmidrule(lr){2-3}\cmidrule(lr){4-7}
        kNN-VC         & -        & -       & 23.65               & 3.93                & 2.24             & 0.66               \\
        SKQ            & c        & c       &  2.25               & 6.88                & 3.03             & 0.77               \\
        SKQ+$\sigma$   & c        & c       &  2.57               & 6.15                & 2.92             & 0.76               \\
        SKQ2+$\sigma$  & q        & c       &  2.75               & 5.61                & 2.96             & 0.76               \\
        SKQ3+$\sigma$  & q        & q       &  2.62               & 5.62                & 2.92             & 0.76               \\ \midrule
        Ground truth   & -        & -       &  1.92               & 1.13                & 3.23             & {-}                \\ \bottomrule
    \end{tabular}
\end{table*}
\begin{table*}[t]
    \caption{Ablation study on speaker vector quantization for one-shot VC.}
    \label{table:ablation}
    \centering
     \sisetup{
    detect-weight, %
    mode=text, %
    tight-spacing=true,
    round-mode=places,
    round-precision=2,
    table-format=1.2,
    table-number-alignment=center
    }
    \begin{tabular}{l*{2}{S[table-format=2.0,round-precision=0]}S[table-format=2.2]SSS}
        \toprule
        Model          & {\#group} & {\#RVQ layer}& {WER [\%] $\downarrow$} & {SV [\%]  $\downarrow$} & {UTMOSv2 $\uparrow$} & {$F_0$-PCC $\uparrow$} \\  
        \cmidrule(lr){1-1}\cmidrule(lr){2-3}\cmidrule(lr){4-7}
        SKQ2+$\sigma$  & {\phantom{0}-}       & {\phantom{0}-}          & 2.75                & 5.61                & 2.96             & 0.76               \\
        SKQ3+$\sigma$  & 4       & 4          & 2.61                & 12.13               & 3.02             & 0.76               \\
        SKQ3+$\sigma$  & 8       & 8          & 2.45                & 9.68                & 3.07             & 0.76               \\
        SKQ3+$\sigma$  & 16      & 8          & 2.62                & 5.62                & 2.92             & 0.76               \\
        SKQ3+$\sigma$  & 16      & 16         & 2.58                & 7.47                & 2.99             & 0.76               \\ \bottomrule
    \end{tabular}
\end{table*}
\subsection{Experimental conditions}
\label{subsec:conditions}
We compare \revise{the proposed method with the following unsupervised methods}:
\begin{itemize}
    \item \textbf{\revise{Vanilla} RVQ}: 
        A low bit-rate variant of DAC~\cite{kumar2023highfidelity}. 
        The encoder consists of downsampling convolutional blocks with strides, 
        where the downsampling factors are set to (2, 4, 5, 8), 
        resulting in a token rate of 50 Hz for 16 kHz audio, which matches the sampling \revise{frame} rate used in WavLM.
        Following the encoder, a two-layer RVQ module with each codebook of size 4096 is applied. 
        Unlike~\cite{kumar2023highfidelity}, depth-wise convolution is incorporated into each convolutional layer, 
        which is essential for stable training in low bit-rate scenarios~\cite{siuzdak2024snac}.
    \item \textbf{SNAC}~\cite{siuzdak2024snac}:
        This model employs the same encoder architecture as \revise{Vanilla} RVQ but utilizes 
        a three-layer RVQ with each codebook of size 4096 and strides of (4, 2, 1).  
    \item \textbf{SKQ}~\cite{sim2024skqvc}: 
        The starting point for this study, it is the same as the architecture described in Section~\ref{sec:prop}, except for the following. 
        The operation IN is simplified to only estimating and subtracting the mean $\bs{\mu(R)}$. 
        Also, neither $\bs{\mu(R)}$, nor the projection of $\mathrm{IN(\mb{R})}$ to an 8-dimensional feature space are quantized by GRVQ and RVQ, respectively, meaning the speaker and prosody information remain continuous. 
        Finally, for decoding, the FiLM operation is replaced by adding $\bs{\mu(R)}$ element-wise to the concatenated content and prosody vectors. Equivalently, this corresponds to $\mb{S}\!=\!\bs{\mu(R)}$, $f(\mb{S})\!=\!1$ and $h(\mb{S})\!=\!\mb{S}$ in \eqref{eq:film}.
    \item \textbf{SKQ+$\sigma$}:
        Same as the architecture described in Section~\ref{sec:prop}, except that neither $\bs{\mu(R)}$ and $\bs{\sigma(R)}$ nor the projection of $\mathrm{IN(\mb{R})}$ to an 8-dimensional feature space are quantized by GRVQ and RVQ, respectively, meaning the speaker and prosody information remain continuous.
    \item \textbf{SKQ2+$\sigma$}:
        Same as the architecture described in Section~\ref{sec:prop}, except that $\bs{\mu(R)}$ and $\bs{\sigma(R)}$ are not quantized by GRVQ, meaning the speaker information remains continuous. 
        $\mathrm{IN(\mb{R})}$ is projected to an 8-dimensional feature space before being quantized by a two-layer RVQ, where each VQ layer has a codebook size of 1000.
    \item \textbf{SKQ3+$\sigma$}:
        The architecture described in Section~\ref{sec:prop}. $\mathrm{IN(\mb{R})}$ is projected to an 8-dimensional feature space before being quantized by a two-layer RVQ, where each VQ layer has a codebook size of 1000. $\bs{\mu(R)}$ and $\bs{\sigma(R)}$ are quantized by GRVQ, where codebook size for each GRVQ layer is fixed at 1024. 
        Other GRVQ hyperparameters are discussed in the ablation study of Section~\ref{subsec:reconstruction}.
    \item \textbf{kNN-VC}~\cite{baas2023voice}: 
        The 6th layer of WavLM is used as a feature extractor. 
        $k$NN regression is performed between the source and target utterances, 
        replacing the matched source features with those from the target.  
\end{itemize}
With the exception of kNN-VC, 
all models are trained on the publicly available LibriSpeech-train-clean-100 dataset and validated on its corresponding validation set, 
using a sampling rate of 16 kHz. 
We train with a batch size of 32 using excerpts of duration 3 seconds,
and validate using excerpts of duration 5 seconds.
The AdamW optimizer is employed with a learning rate of $6\times 10^{-4}$, $\beta_1 = 0.8$, and $\beta_2 = 0.9$.
The learning rate is decayed at each step with a decay factor of $\gamma = 0.999994$.
The weights for the multiple loss functions are identical to~\cite{kumar2023highfidelity, siuzdak2024snac},
and layer dropout is not applied for RVQ and GRVQ.

The pre-trained WavLM-Large model is obtained from its official repository\footnote{https://huggingface.co/microsoft/wavlm-large},
and extracts 1024-dimensional features at 50 Hz for 16 kHz audio from its 6th layer.
Codebooks\footnote{https://huggingface.co/speechbrain/SSL\_Quantization} for WavLM-Large are constructed using mini-batch $k$-means clustering over the full training subset of the LibriSpeech dataset, 
with the number of centroids fixed at 1000.

To ensure a fair comparison, all models except kNN-VC employ the same decoder architecture,
introduced at the end of Section~\ref{subsec:Decoder}.
The upsampling factors of the decoder's upsampling convolutional blocks are set to (8, 5, 4, 2). 
The discriminator follows the architecture proposed in DAC~\cite{kumar2023highfidelity}.
For kNN-VC, the decoder utilizes HiFi-GAN~\cite{kong2020hifigan}, 
which is trained to estimate waveforms from the WavLM-Large 6th-layer representations\footnote{https://github.com/bshall/knn-vc}.

We conduct two types of evaluation, both performed on the LibriSpeech-test-clean subset.
For reconstruction evaluation, we employ the following objective metrics: 
mel-spectrum distance (mel-d), STFT distance (STFT-d), and  ViSQOL\footnote{https://github.com/google/visqol}~\cite{chinen2020visqolv3} for sound quality, and equal error rate (EER) for speaker verification (SV). 
For SV, the speaker embedding is estimated by ECAPA-TDNN~\cite{dawalatabad2021ECAPATDNN}, implemented using the NeMo toolkit\footnote{https://github.com/NVIDIA/NeMo}~\cite{kuchaiev2019nemo}.
For one-shot VC evaluation,
\revise{we randomly select two reference utterances: one spoken by a male speaker and the other by a female speaker,
and speaker identity is modified by swapping the quantized speaker vector with that of a reference utterance.}
We assess the word error rate (WER) of the converted speech, EER of SV, 
UTMOSv2 scores\footnote{https://github.com/sarulab-speech/UTMOSv2}~\cite{baba2024t05system},
and the Pearson correlation coefficient (PCC) of $F_0$ between the $F_0$ contours of the source and converted speech. 
WER is calculated using ASR with Whisper-turbo\footnote{https://github.com/openai/whisper}~\cite{radford2022robust}.
For $F_0$-PCC, %
$F_0$ estimates are obtained using
DIO~\cite{morise2009fast} and StoneMask~\cite{morise2009WORLD} from the WORLD vocoder\footnote{https://github.com/JeremyCCHsu/Python-Wrapper-for-World-Vocoder}.

\begin{figure}[t]
    \begin{subfigure}{0.45\linewidth}
        \framebox{\includegraphics[width=.99\linewidth]{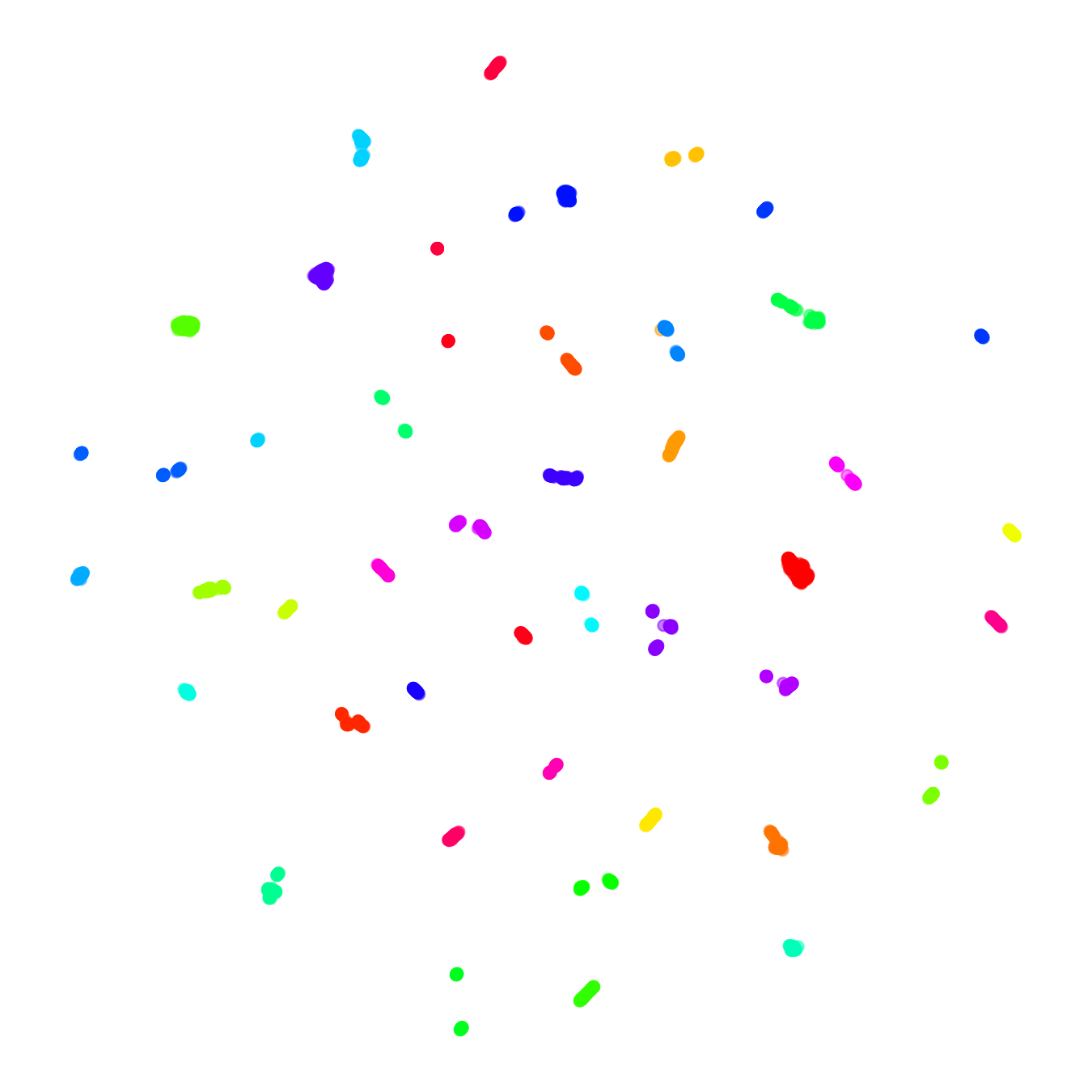}}
        \caption{ECAPA-TDNN}
    \end{subfigure}
     \hspace{.3cm}
    \begin{subfigure}{0.45\linewidth}
        \framebox{\includegraphics[width=.99\linewidth]{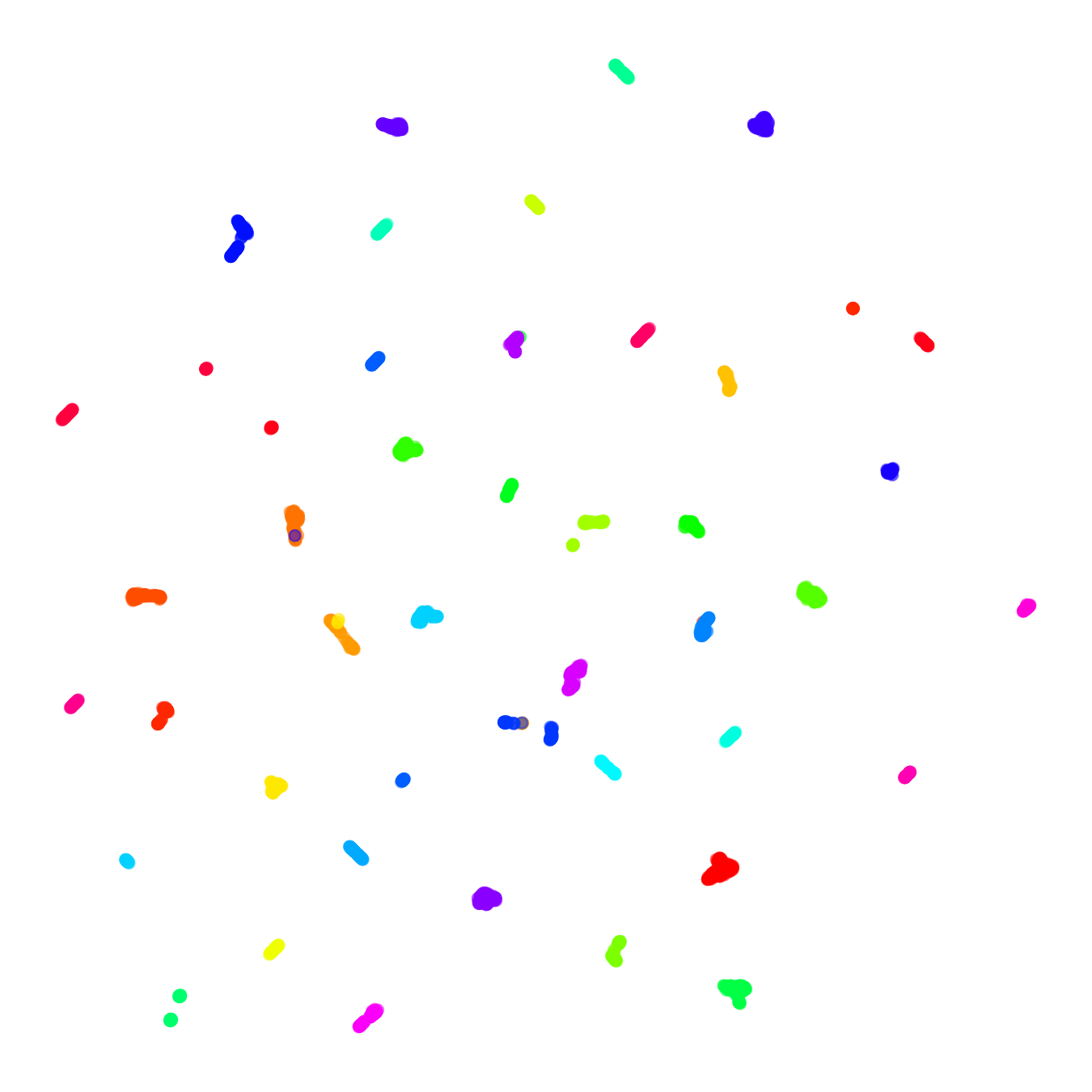}}
        \caption{$\mb{S}$ from SKQ3+$\sigma$}
    \end{subfigure}
    
    \begin{subfigure}{0.45\linewidth}
    \vspace{.3cm}
        \framebox{\includegraphics[width=.99\linewidth]{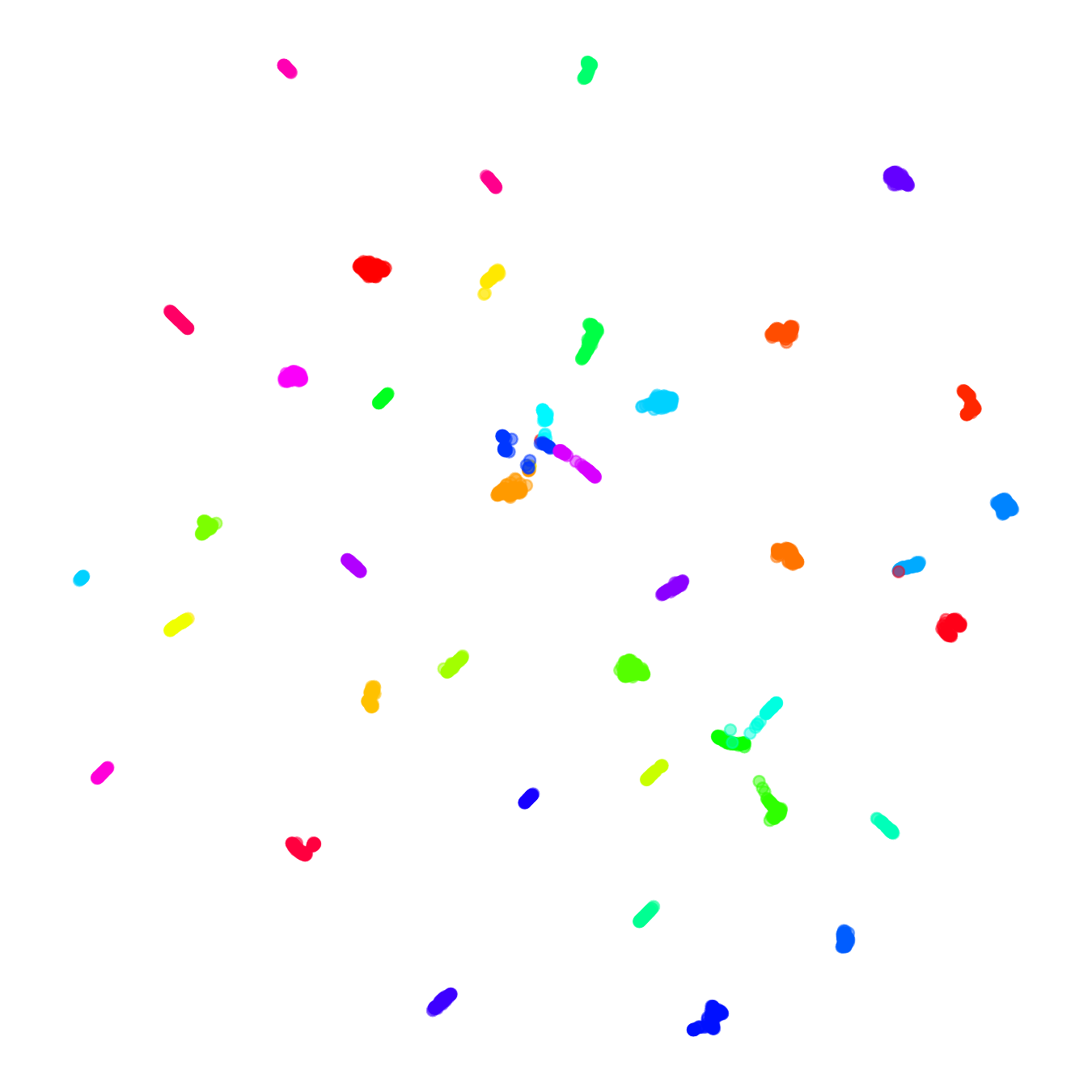}}
        \vspace{-0.31cm} 
        \caption{$\mb{W}$ from WavLM}
    \end{subfigure}
    \hspace{.3cm}
    \begin{subfigure}{0.45\linewidth}
        \framebox{\includegraphics[width=.99\linewidth]{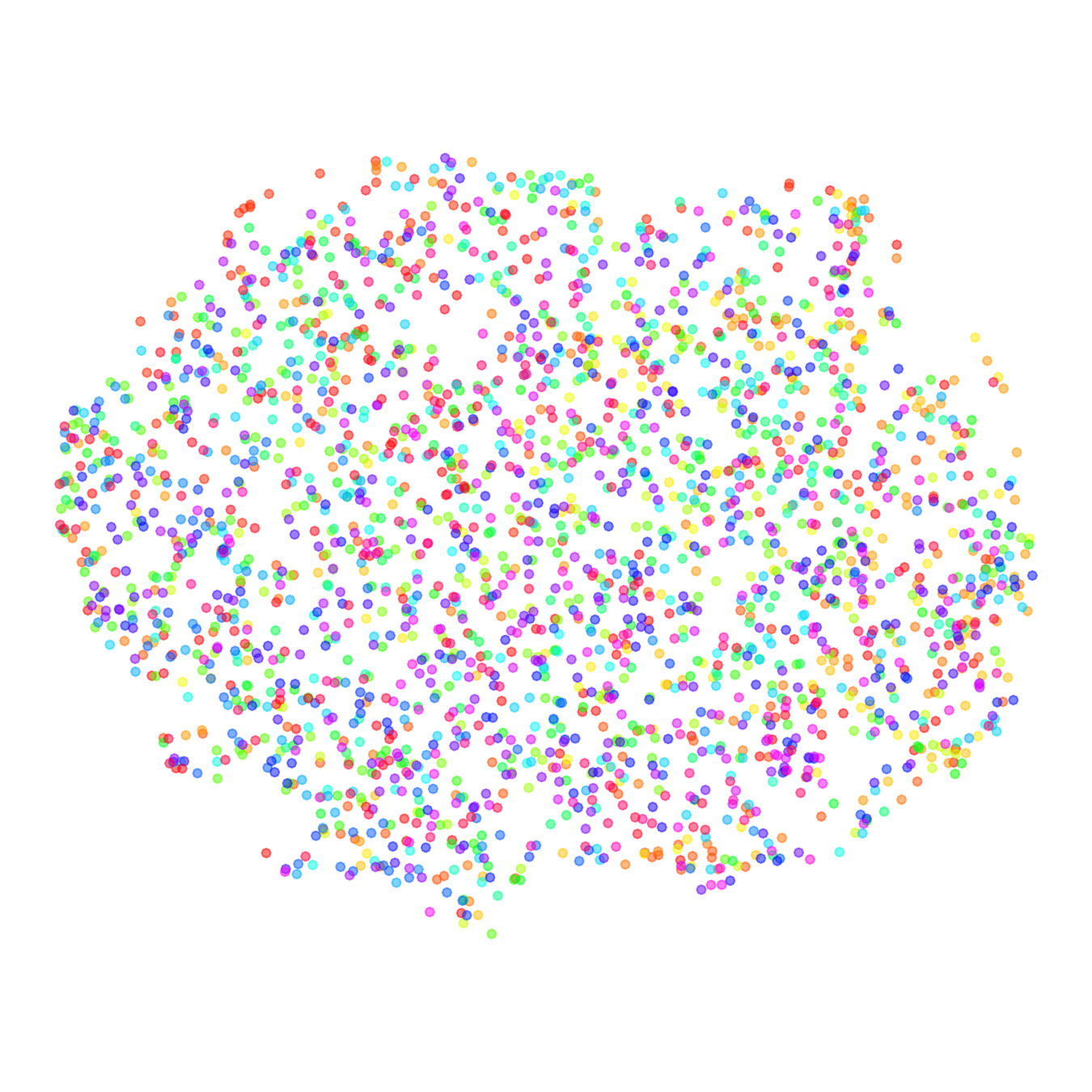}}
        \vspace{-0.31cm} 
        \caption{$\mb{P}$ from SKQ3+$\sigma$}
    \end{subfigure}
    \caption{UMAPs for LibriSpeech utterance. Colors represent distinct speakers.}
    \vspace{-0.3cm} 
    \label{fig:emb}
\end{figure}

\begin{figure}[t]
    \centering
    \framebox{\includegraphics[width=0.8\linewidth]{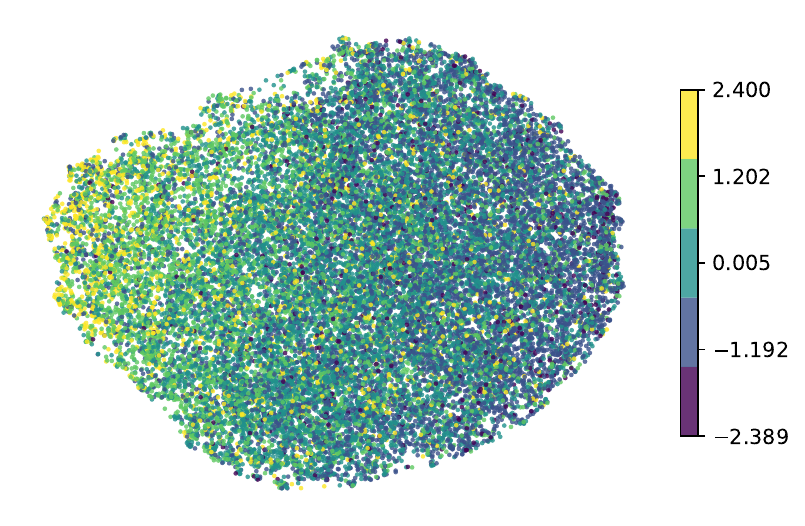}}
    \caption{UMAP of $\mb{P}$ from SKQ3+$\sigma$. Colors represent $F_{0}$ normalized by each speaker’s mean and variance.}
    \label{fig:F0emb}
\end{figure}

\begin{figure}[t]
  \centering
    \includegraphics[width=0.48\textwidth]{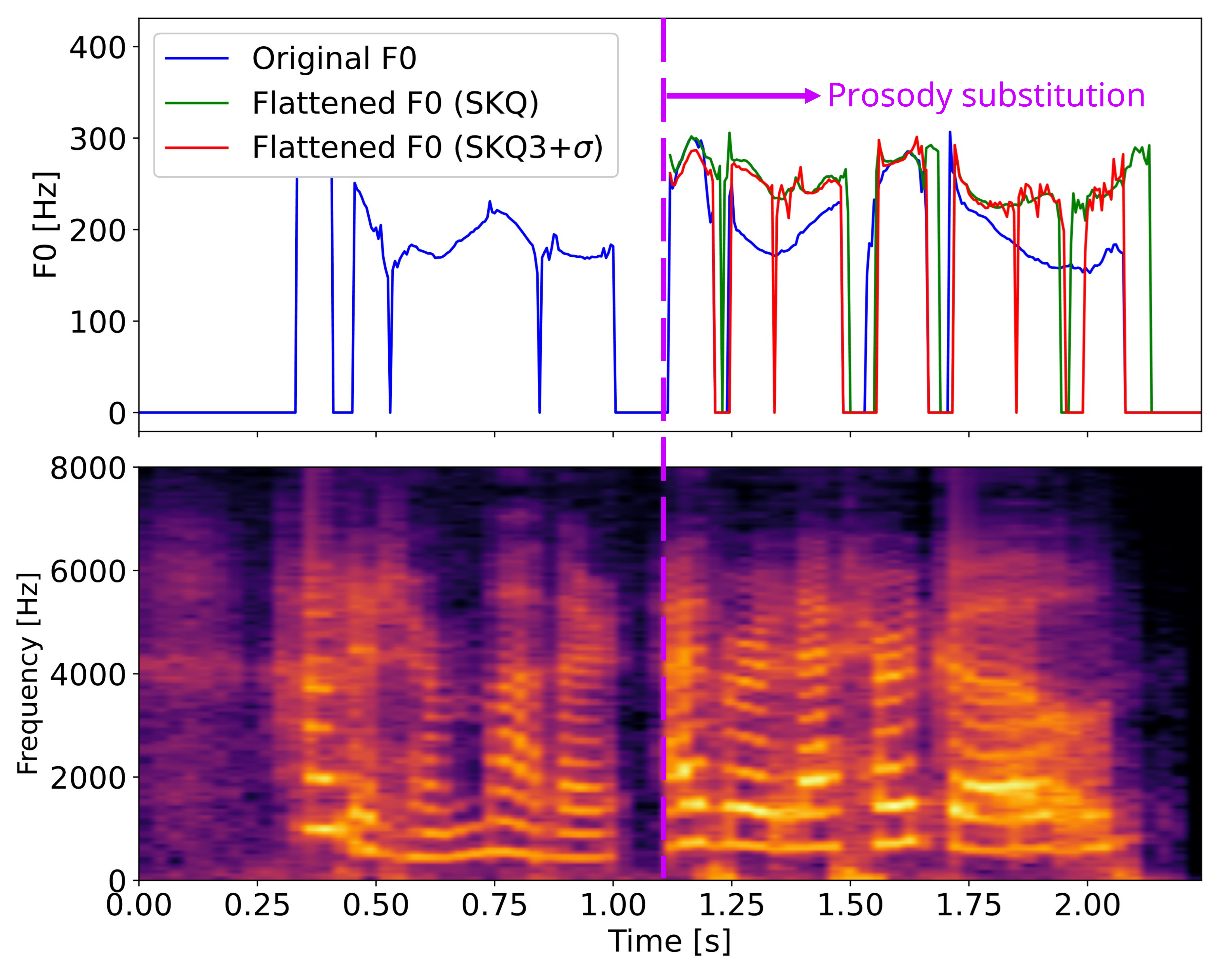}
  \caption{$F_0$ contours of the original and flattened samples, along with the $F_0$-flattened spectrogram (SKQ3+$\sigma$).} 
  \label{fig:spectrogram_f0}
\end{figure}
\subsection{Experimental results}
\label{subsec:reconstruction}
Table~\ref{table:reconstruction} presents the reconstruction evaluation results,
where the values for phonetic, prosodic, and speaker indicate the amount of data associated with each (continuous or quantized) vector representing the corresponding information, and
kb/s, kb/u, and dim/u stand for kilobits per second, kilobits per utterance (for quantized vectors), and dimension per utterance (for continuous vectors), respectively.
The non-disentangled methods, Vanilla RVQ and SNAC, 
achieve slightly higher reconstruction quality than disentangled methods. 
However, disentangled methods demonstrate superior SV accuracy.
Quantizing both prosody and speaker vectors results in a slight degradation of SV and ViSQOL performance, 
indicating some loss of information when converting to an LLM-compatible discrete representation.

Table~\ref{table:conversion} presents the one-shot VC evaluation results,
where ``c'' and ``q'' indicate continuous and quantized vectors, respectively.
The conventional kNN-VC achieved the highest SV accuracy but performed the worst across other evaluation metrics.
This outcome is likely due to the converted speech being generated solely from the 6th-layer representation of WavLM for the target speaker.
Incorporating the variance component of the speaker vector improves SV accuracy.
Quantizing the prosody vectors further enhances SV performance, 
presumably by introducing an information bottleneck that facilitates the disentanglement of prosodic and phonetic information.
However, quantizing speaker vectors slightly degrades SV performance, 
posing a trade-off in establishing compatibility with LLMs, while other evaluation metrics remain unaffected.

Table~\ref{table:ablation} presents the results of the ablation study on speaker vector quantization in one-shot VC,
where ``\#group'' and ``\#RVQ layer'' denote the number of groups and the number of RVQ layers in the GRVQ block of Fig.~\ref{fig:main}(b), respectively.
These results suggest that increasing the bitrate for speaker vectors may be advantageous for improving SV accuracy \revise{without substantially affecting all the other VC performance indicators}.

Figure~\ref{fig:emb} presents the UMAPs\footnote{https://github.com/lmcinnes/umap}
of (a) source speaker embeddings (obtained from ECAPA-TDNN), 
(b) the quantized speaker vectors $\mb{S}$ (from SKQ3+$\sigma$),
(c) $\mb{W}$ (from WavLM), and (d) $\mb{P}$ (from SKQ3+$\sigma$), respectively.
Here, SKQ3+$\sigma$ has a GRVQ with 16 groups and 8 codebooks. \revise{We average $\mb{W}$ and $\mb{P}$ over time for each utterance.}
Each color represents a distinct speaker. 
The projections indicate that both the quantized speaker vectors and 
$\mb{W}$ from WavLM form clusters corresponding to the identity of the utterance speaker, 
consistent with the behavior of speaker verification models like ECAPA-TDNN.
In contrast, $\mb{P}$ does not exhibit distinct speaker clusters, demonstrating the feature disentanglement between S and P in our proposed model.

\revise{
Figure~\ref{fig:F0emb} shows the UMAP of $\mb{P}$ from SKQ3+$\sigma$,
where the color represents $F_{0}$ of each given frame normalized by the corresponding speaker's mean and variance.
Here, SKQ3+$\sigma$ employs a GRVQ with 16 groups and 8 codebooks.
The projection indicates that quantized prosody vectors align with $F_0$ deviation, with points corresponding to similar amounts of frequency deviation staying together.
}

To illustrate the controllability of prosody in SKQ3+$\sigma$,
we select a sample from the test set and resynthesize it
by replacing the prosody codes $\mb{P}$ in the latter half of the utterance with a fixed value equal to the code at the midpoint,
resulting in a flattened prosody
while preserving some variation.
The spectrogram of the generated speech and the $F_0$ contours of the original and generated speech are shown in Fig.~\ref{fig:spectrogram_f0},
where the prosody in the latter half is flattened, maintaining both naturalness and speaker characteristics.
\revise{The ability of SKQ3+$\sigma$ to perform this adjustment on quantized embeddings enables fine-grained $F_0$ (i.e., disentangled prosodic) control in models that require discrete codes, i.e., LLM-compatible models~\cite{kyutai2024moshi}.}
\section{Conclusion and future works}
\label{sec:conc}
This paper explored a neural speech codec designed to proactively disentangle phonetic, prosodic, and speaker characteristics 
without 
supervision from textual or speaker information,
leveraging the strong correlation between speaker identity and the middle layers of an SSL model. 
\revise{Unlike conventional VC methods, 
not only the phonetic feature but also prosodic and 
speaker features are also quantized, 
which facilitates the integration of these features into LLMs.}
Experimental results demonstrate that the proposed disentangled speech codec does effectively generate disentangled representations while preserving reconstruction quality comparable 
to its non-disentangled counterpart, and 
achieving performance on par with conventional VC methods. 
An ablation study further indicates that expressivity of the quantized representation of the speaker vector 
impacts speaker identity control, underscoring the necessity of allocating sufficient bitrate for sufficient speaker representation.

\revise{
In future work, we will test the learned disentangled representations
in relevant downstream tasks (e.g., ASR, TTS). 
We will also explore representing speech with discrete codec tokens 
in speech LMs, enabling the prediction of phonetic and 
prosodic sequences together with 
a fixed speaker code to maintain speaker consistency 
in duplex dialogue systems.
}
\clearpage

\bibliographystyle{IEEEtran}
\bibliography{codec}

\begin{thebibliography}{10}
\providecommand{\url}[1]{#1}
\csname url@samestyle\endcsname
\providecommand{\newblock}{\relax}
\providecommand{\bibinfo}[2]{#2}
\providecommand{\BIBentrySTDinterwordspacing}{\spaceskip=0pt\relax}
\providecommand{\BIBentryALTinterwordstretchfactor}{4}
\providecommand{\BIBentryALTinterwordspacing}{\spaceskip=\fontdimen2\font plus
\BIBentryALTinterwordstretchfactor\fontdimen3\font minus \fontdimen4\font\relax}
\providecommand{\BIBforeignlanguage}[2]{{%
\expandafter\ifx\csname l@#1\endcsname\relax
\typeout{** WARNING: IEEEtran.bst: No hyphenation pattern has been}%
\typeout{** loaded for the language `#1'. Using the pattern for}%
\typeout{** the default language instead.}%
\else
\language=\csname l@#1\endcsname
\fi
#2}}
\providecommand{\BIBdecl}{\relax}
\BIBdecl

\bibitem{borsos2023audiolm}
Z.~Borsos, R.~Marinier, D.~Vincent, E.~Kharitonov, O.~Pietquin, M.~Sharifi, D.~Roblek, O.~Teboul, D.~Grangier, M.~Tagliasacchi \emph{et~al.}, ``{AudioLM}: a language modeling approach to audio generation,'' \emph{IEEE/ACM Trans. Audio, Speech, Lang. Process.}, vol.~31, pp. 2523--2533, 2023.

\bibitem{kyutai2024moshi}
A.~D\'efossez, L.~Mazar\'e, M.~Orsini, A.~Royer, P.~P\'erez, H.~J\'egou, E.~Grave, and N.~Zeghidour, ``Moshi: A speech-text foundation model for real-time dialogue,'' \emph{arXiv preprint arXiv:2410.00037}, 2024.

\bibitem{mousavi2024dasb}
P.~Mousavi, L.~{Della Libera}, J.~Duret, A.~Ploujnikov, C.~Subakan, and M.~Ravanelli, ``{DASB} - discrete audio and speech benchmark,'' \emph{arXiv preprint arXiv:2406.14294}, 2024.

\bibitem{zeghidour2021soundstream}
N.~Zeghidour, A.~Luebs, A.~Omran, J.~Skoglund, and M.~Tagliasacchi, ``{SoundStream}: An end-to-end neural audio codec,'' \emph{IEEE/ACM Trans. Audio, Speech, Lang. Process.}, vol.~30, pp. 495--507, 2021.

\bibitem{defossez2022highfi}
A.~D{\'e}fossez, J.~Copet, G.~Synnaeve, and Y.~Adi, ``High fidelity neural audio compression,'' \emph{TMLR}, 2023.

\bibitem{kumar2023highfidelity}
R.~Kumar, P.~Seetharaman, A.~Luebs, I.~Kumar, and K.~Kumar, ``High-fidelity audio compression with improved {RVQGAN},'' in \emph{Proc. NeurIPS}, 2023.

\bibitem{zhang2024speechtokenizer}
X.~Zhang, D.~Zhang, S.~Li, Y.~Zhou, and X.~Qiu, ``{SpeechTokenizer}: Unified speech tokenizer for speech large language models,'' in \emph{Proc. ICLR}, 2024.

\bibitem{huang2024repcodec}
Z.~Huang, C.~Meng, and T.~Ko, ``{RepCodec}: A speech representation codec for speech tokenization,'' in \emph{Proc. ACL}, 2024.

\bibitem{ye2024codec}
Z.~Ye, P.~Sun, J.~Lei, H.~Lin, X.~Tan, Z.~Dai, Q.~Kong, J.~Chen, J.~Pan, Q.~Liu, Y.~Guo, and W.~Xue, ``Codec does matter: Exploring the semantic shortcoming of codec for audio language model,'' \emph{arXiv preprint arXiv:2408.17175}, 2024.

\bibitem{fujisaki2003prosody}
H.~Fujisaki, ``Prosody, information, and modeling — with emphasis on tonal features of speech —,'' in \emph{Proc. Interspeech}, 2003.

\bibitem{baas2023voice}
M.~Baas, B.~van Niekerk, and H.~Kamper, ``Voice conversion with just nearest neighbors,'' in \emph{Proc. Interspeech}, 2023.

\bibitem{chen2022wavlm}
S.~Chen, C.~Wang, Z.~Chen, Y.~Wu, S.~Liu, Z.~Chen, J.~Li, N.~Kanda, T.~Yoshioka, X.~Xiao, J.~Wu, L.~Zhou, S.~Ren, Y.~Qian, Y.~Qian, J.~Wu, M.~Zeng, X.~Yu, and F.~Wei, ``{WavLM}: Large-scale self-supervised pre-training for full stack speech processing,'' \emph{IEEE J. Sel. Top. Signal Process.}, vol.~16, no.~6, pp. 1505--–1518, 2022.

\bibitem{hajal2025knn}
K.~E. Hajal, A.~Kulkarni, E.~Hermann, and M.~Magimai-Doss, ``{kNN} retrieval for simple and effective zero-shot multi-speaker text-to-speech,'' in \emph{Proc. NAACL}, 2025.

\bibitem{sim2024skqvc}
Y.~Sim, J.~Yoon, and Y.~Suh, ``{SKQVC}: One-shot voice conversion by k-means quantization with self-supervised speech representations,'' \emph{arXiv preprint arXiv:2411.16147}, 2024.

\bibitem{guo2024lscodec}
Y.~Guo, Z.~Li, C.~Du, H.~Wang, X.~Chen, and K.~Yu, ``{LSCodec}: Low-bitrate and speaker-decoupled discrete speech codec,'' \emph{arXiv preprint arXiv:2410.15764}, 2024.

\bibitem{qian2022contentvec}
K.~Qian, Y.~Zhang, H.~Gao, J.~Ni, C.~Lai, D.~Cox, M.~Hasegawa-Johnson, and S.~Chang, ``{ContentVec}: An improved self-supervised speech representation by disentangling speakers,'' \emph{arXiv preprint arXiv:2204.09224}, 2022.

\bibitem{ren2024fewertoken}
Y.~Ren, T.~Wang, J.~Yi, L.~Xu, J.~Tao, C.~Zhang, and J.~Zhou, ``Fewer-token neural speech codec with time-invariant codes,'' in \emph{Proc. ICASSP}, 2024.

\bibitem{dellalibera2025focalcodecl}
L.~{Della Libera}, F.~Paissan, C.~Subakan, and M.~Ravanelli, ``{FocalCodec}: Low-bitrate speech coding via focal modulation networks,'' \emph{arXiv preprint arXiv:2502.04465}, 2025.

\bibitem{wang2021vqmivc}
D.~Wang, L.~Deng, Y.~Yeung, X.~Chen, X.~Liu, and H.~Meng, ``{VQMIVC}: Vector quantization and mutual information-based unsupervised speech representation disentanglement for one-shot voice conversion,'' in \emph{Proc. Interspeech}, 2021.

\bibitem{kanagawa2024knowledgs}
H.~Kanagawa and Y.~Ijima, ``Knowledge distillation from self-supervised representation learning model with discrete units for any-to-any streaming voice conversion,'' in \emph{Proc. Interspeech}, 2024.

\bibitem{polyak2021speechresynthesis}
A.~Polyak, Y.~Adi, J.~Copet, E.~Kharitonov, K.~Lakhotia, W.-N. Hsu, A.~Mohamed, and E.~Dupoux, ``Speech resynthesis from discrete disentangled self-supervised representations,'' in \emph{Proc. Interspeech}, 2021.

\bibitem{karapiperis2024investigating}
S.~Karapiperis, N.~Ellinas, A.~Vioni, J.~Oh, G.~Jho, I.~Hwang, and S.~Raptis, ``Investigating disentanglement in a phoneme-level speech codec for prosody modeling,'' in \emph{Proc. SLT}, 2024.

\bibitem{ju2024naturalspeech}
Z.~Ju, Y.~Wang, K.~Shen, X.~Tan, D.~Xin, D.~Yang, Y.~Liu, Y.~Leng, K.~Song, S.~Tang, Z.~Wu, T.~Qin, X.-Y. Li, W.~Ye, S.~Zhang, J.~Bian, L.~He, J.~Li, and S.~Zhao, ``{NaturalSpeech} 3: Zero-shot speech synthesis with factorized codec and diffusion models,'' \emph{arXiv preprint arXiv:2403.03100}, 2024.

\bibitem{siuzdak2024snac}
H.~Siuzdak, F.~Gr{\"o}tschla, and L.~A. Lanzend{\"o}rfer, ``{SNAC}: Multi-scale neural audio codec,'' in \emph{Proc. Audio Imagination: NeurIPS Workshop on AI-Driven Speech, Music, and Sound Generation}, 2024.

\bibitem{series2020adversarial}
H.~Serieys and A.~Cherif, ``Adversarial domain adaptation without gradient reversal layer,'' in \emph{Proc. ESANN}, 2020.

\bibitem{kong2020hifigan}
J.~Kong, J.~Kim, and J.~Bae, ``{HiFi-GAN}: Generative adversarial networks for efficient and high fidelity speech synthesis,'' in \emph{Proc. NeurIPS}, 2020.

\bibitem{jang2021univnet}
W.~Jang, D.~Lim, J.~Yoon, B.~Kim, and J.~Kim, ``{UnivNet}: A neural vocoder with multi-resolution spectrogram discriminators for high-fidelity waveform generation,'' in \emph{Proc. Interspeech}, 2021.

\bibitem{mousavi2024how}
P.~Mousavi, J.~Duret, S.~Zaiem, L.~Della~Libera, A.~Ploujnikov, C.~Subakan, and M.~Ravanelli, ``How should we extract discrete audio tokens from self-supervised models?'' in \emph{Proc. Interspeech}, 2024.

\bibitem{hsu2021hubert}
W.-N. Hsu, B.~Bolte, Y.-H.~H. Tsai, K.~Lakhotia, R.~Salakhutdinov, and A.~Mohamed, ``{HuBERT}: Self-supervised speech representation learning by masked prediction of hidden units,'' \emph{IEEE/ACM Trans. Audio, Speech, Lang. Process.}, vol.~29, pp. 3451 -- 3460, 2021.

\bibitem{chen2020againvc}
Y.~Chen, D.~Wu, T.~Wu, and H.~Lee, ``{AGAIN-VC}: A one-shot voice conversion using activation guidance and adaptive instance normalization,'' in \emph{Proc. ICASSP}, 2021.

\bibitem{yang2023hifi}
D.~Yang, S.~Liu, R.~Huang, J.~Tian, C.~Weng, and Y.~Zou, ``{HiFi-Codec}: Group-residual vector quantization for high fidelity audio codec,'' \emph{arXiv preprint arXiv:2305.02765}, 2023.

\bibitem{perez2018film}
E.~Perez, F.~Strub, H.~de~Vries, V.~Dumoulin, and A.~Courville, ``{FiLM}: Visual reasoning with a general conditioning layer,'' in \emph{Proc. AAAI}, 2018.

\bibitem{chinen2020visqolv3}
M.~Chinen, F.~S.~C. Lim, J.~Skoglund, N.~Gureev, F.~O'Gorman, and A.~Hines, ``{ViSQOL} v3: An open source production ready objective speech and audio metric,'' in \emph{Proc. QoMEX}, 2020.

\bibitem{dawalatabad2021ECAPATDNN}
N.~Dawalatabad, M.~Ravanelli, F.~Grondin, J.~Thienpondt, B.~Desplanques, and H.~Na, ``{ECAPA-TDNN} embeddings for speaker diarization,'' in \emph{Proc. Interspeech}, 2021.

\bibitem{kuchaiev2019nemo}
O.~Kuchaiev, J.~Li, H.~Nguyen, O.~Hrinchuk, R.~Leary, B.~Ginsburg, S.~Kriman, S.~Beliaev, V.~Lavrukhin, J.~Cook, P.~Castonguay, M.~Popova, J.~Huang, and J.~M. Cohen, ``{NeMo}: A toolkit for building {AI} applications using neural modules,'' \emph{arXiv preprint arXiv:1909.09577}, 2019.

\bibitem{baba2024t05system}
K.~Baba, W.~Nakata, Y.~Saito, and H.~Saruwatari, ``The {T05} system for the {VoiceMOS} challenge 2024: Transfer learning from deep image classifier to naturalness {MOS} prediction of high-quality synthetic speech,'' in \emph{Proc. SLT}, 2024.

\bibitem{radford2022robust}
A.~Radford, J.~W. Kim, T.~Xu, G.~Brockman, C.~McLeavey, and I.~Sutskever, ``Robust speech recognition via large-scale weak supervision,'' \emph{arXiv preprint arXiv:2212.04356}, 2022.

\bibitem{morise2009fast}
M.~Morise, H.~Kawahara, and H.~Katayose, ``Fast and reliable {F0} estimation method based on the period extraction of vocal fold vibration of singing voice and speech,'' in \emph{Proc. AES Int. Conf.}, 2009.

\bibitem{morise2009WORLD}
M.~Morise, F.~Yokomori, and K.~Ozawa, ``{WORLD}: A vocoder-based high-quality speech synthesis system for real-time applications,'' \emph{IEICE Trans. Inf. \& Syst.}, vol. E99.D, no.~7, pp. 1877--1884, 2016.

\end{thebibliography}

\end{document}